\documentclass{elsart5p}

\usepackage{graphics}
\usepackage{graphicx}
\usepackage{amssymb}
\begin{document}

\begin{frontmatter}



\title{Pomeranchuk instability: symmetry breaking and experimental signatures}
%

\author[1]{J. Quintanilla\corauthref{Name1}}
\ead{j.quintanilla@rl.ac.uk},
\author[2]{C. Hooley},
\author[3]{B. J. Powell},
\author[4]{A. J. Schofield} and
\author[5]{M. Haque}.

\address[1]{ISIS Spallation Facility, 
STFC Rutherford Appleton Laboratory, Harwell Science and Innovation
Campus, Didcot, OX11 0QX, U.K.}
\address[2]{School of Physics and Astronomy, University of St. Andrews, North
Haugh, St. Andrews, Fife, KY16 9SS, U.K.}
\address[3]{Department of
Physics, University of Queensland, Brisbane, Queensland 4072,
Australia}
\address[4]{School of Physics and Astronomy, University of Birmingham,
Edgbaston, Birmingham, B15 2TT, U.K.}
\address[5]{Max Planck Institute for the Physics of Complex Systems, Noethnitzer
Str. 38, 01187 Dresden, Germany}

\corauth[Name1]{Corresponding author. Tel: +44 (0)123 544 6353 fax: +44 (0)123
544 5720}

\begin{abstract}
We discuss the emergence of symmetry-breaking {\it via} the  Pomeranchuk instability from interactions that respect the underlying point-group symmetry. We use a variational mean-field theory to consider a 2D continuum and a square lattice. We describe two experimental signatures:  a symmetry-breaking pattern of Friedel oscillations around an impurity; and a structural transition. 
\end{abstract}


\end{frontmatter}


Through the Pomeranchuk instability (PI)
\cite{1958-Pomeranchuk} a Fermi liquid may enter a ``nematic'' state that breaks rotational symmetries while preserving all translations. This might be a ``missing link'' in the phase diagram of strongly-correlated quantum matter joining the Wigner crystal or Mott insulator, through stripe (smectic) phases to the Fermi liquid \cite{2007-Fradkin-Kivelson-Oganesyan}. One way to describe it phenomenologically is to assume effective interactions that are \emph{anisotropic} (in a two-dimensional continuum \cite{2001-Oganesyan-Kivelson-Fradkin}) or that break the symmetry of the crystal (on a square lattice \cite{2004-Khavkine-Chung-Oganesyan-Kee}). More microscopically, it has been shown that rotational symmetry-breaking can emerge in three dimensions from \emph{isotropic} interactions \cite{2006-Quintanilla-Schofield}. Here we show that a similar argument applies in two dimensions and to the square lattice. We will then mention two possible experimental signatures of the PI that arise in these simple contexts. These may be relevant to quantum Hall systems \cite{2007-Doan-Manousakis}, URu$_2$Si$_2$ \cite{2006-Varma-Zhu} and Sr$_3$Ru$_2$O$_7$ \cite{Sr327}.

In a two-dimensional continuum, we proceed in a way entirely analogous to Ref.~\cite{2006-Quintanilla-Schofield}. Our starting point is a many-body Hamiltonian featuring free fermions, of mass $m$ each, and an \emph{isotropic}, spin-independent interaction potential $V(|{\bf r}-{\bf r}'|)$. Its Fourier transform $V({\bf K})\equiv \int d^2{\bf R} e^{-i{\bf K}.{\bf R}}V(|{\bf R}|)$ (with dimensions ${\rm energy} \times {\rm area}$) contains terms in all angular momentum channels:
\begin{equation}
  V({\bf k}-{\bf k}')=\sum_{l=0}^{\infty} V_l(|{\bf k}|,|{\bf k}'|)
  \cos(l(\theta_{\bf k}-\theta_{{\bf k}'})).
  \label{V-expansion}
\end{equation}
We try a variational ground state $\left| \Psi \right\rangle = \prod_{\epsilon({\bf k})<0} \hat{c}^{\dagger}_{{\bf k},\sigma}\left| 0 \right\rangle$ featuring a renormalised dispersion relation $\epsilon({\bf k})$ which we treat as the variational parameter. To find an instability equation, let us assume that it is made up of an isotropic component plus a small, symmetry-breaking contribution. Near the instability, the latter must have a well-defined angular momentum quantum number $l=1,2,3,4,\ldots$ Without loss of generality we write
\(
  \epsilon({\bf k}) = \epsilon_0(|{\bf k}|) + \Lambda_l(|{\bf k}|)\cos(l\theta_{\bf k})
\)
and find the following self-consistency equation:
\begin{equation}
  \Lambda_l(k)=\frac{1}{4\pi^2}\int_{0}^{2\pi}
  d\theta \cos(l\theta)\int_{k_F}^{k_F+\delta k_F(\theta)}
  dq ~ q ~ V_l(k,q).
  \label{sceq}
\end{equation}
Here $k_F$ is the ``unperturbed'' Fermi vector, given by $\epsilon_0(k_F)=0,$ and $\delta k_F$ is a small anisotropic distortion, depending on $\Lambda_l(k)$ and $\theta$ through $\epsilon(\left[k_F+\delta k_F\right]\hat{\bf k})=0$. Note that only the $l^{\rm th}$ term of the expansion (\ref{V-expansion}) comes into play near the instability, even though all the other terms are non-zero. Using $\delta k_F \ll k_F$, Eq.~(\ref{sceq}) gives
\begin{equation}
  V_{\rm crit} = \frac{4\pi \hbar v_F}{k_F},
\end{equation}
giving the critical value of the coupling constant $V \equiv V_l(k_F,k_F)$. For $V>V_{\rm crit}$, the distortion is given by 
\(
  \delta k_F(\theta_{\bf k}) = \left(\Lambda/\hbar v_F\right) \cos(l\theta_{\bf k}),
\)
where we have written $\epsilon_0(k_F+\delta k_F) \approx \hbar v_F \delta k_F$ and $\Lambda(k_F+\delta k_F) \approx \Lambda + \Lambda' \delta k_F$ and assumed $\Lambda' \ll \hbar v_F$. It grows according to the mean field law
\begin{equation}
  \Lambda = \sqrt{\frac{16\pi \hbar^3 v_F^3}{-V' V_{\rm crit}}}\left(V-V_{\rm crit}\right)^{1/2}.
\end{equation}
This is valid only for 
\(
  V' \equiv \partial V(k_F,q) / \partial q|_{q=k_F} < 0.
\)
When $V'>0$, we have a first-order phase transition.

\begin{figure}[!ht]
\begin{center}
\includegraphics[angle=0,width=0.45\textwidth]{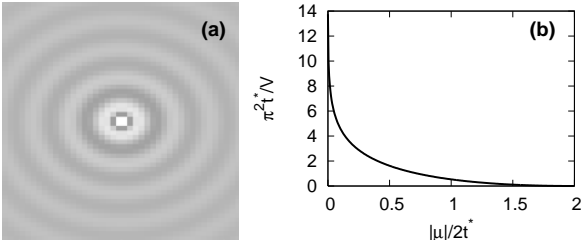}
\end{center}
\caption{(a) LDOS around an impurity for an elliptical Fermi surface with $k_F$ 10 percent larger in the $x$ direction. (b) Ground-state phase diagram for the Pomeranchuk instability with $d_{x^2-y^2}$ symmetry on a square lattice with nearest-neighbour repulsion.} \label{fig.1}
\end{figure}

In the Pomeranchuk state, the homogeneous electron fluid remains uniform. The symmetry breaking is revealed in real space, though, by Friedel oscillations around a single impurity. These could be observed by a scanning tunnelling microscope (STM), which measures the local density of states (LDOS), $D(\omega,{\bf r}) = -\frac{1}{\pi} {\rm Im} G^R ({\bf r},{\bf r};\omega)$ \cite{2003-Fiete-Heller}. For a single, delta-function impurity at the origin we have 
\begin{equation}
  G^R({\bf r},{\bf r}')=G_0^R({\bf r},{\bf r}';\omega)
  + \Gamma G_0^R({\bf r},0;\omega)G_0^R(0,{\bf r}';\omega)
\end{equation}
where $\Gamma$ is a renormalized coupling constant and $G_0^R$ is the retarded Green's function in the absence of the impurity (the details will be published elsewhere \cite{Schofield-Hooley-Quintanilla}). Fig.~\ref{fig.1}~(a) shows the resulting pattern of Friedel oscillations, from which the broken symmetry is apparent.

The above arguments on symmetry breaking can be carried over to crystal lattices. Consider spin-$1/2$ fermions on a square lattice with nearest-neighbour hopping and repulsion on-site (with strength $U$) and between nearest neighbours (with strength $V$): \cite{2001-Valenzuela-Vozmediano} 
\begin{eqnarray}
\hat{H} &=& -t\sum_{\alpha=\uparrow,\downarrow}
\sum_{<i,j>} \hat{c}_{i,\alpha}^{\dagger} \hat{c}_{j,\alpha}
+\sum_{\alpha=\uparrow,\downarrow}\sum_{i} \frac{U }{2} \hat{c}_{i,\alpha}^{\dagger} \hat{c}_{i,-\alpha}^{\dagger}
\hat{c}_{i,-\alpha} \hat{c}_{i,\alpha}
\nonumber
\\&&
+\sum_{\alpha,\beta=\uparrow,\downarrow}\sum_{<i,j>} \frac{V}{2} \hat{c}_{i,\alpha}^{\dagger} \hat{c}_{j,\beta}^{\dagger}
\hat{c}_{j,\beta} \hat{c}_{i,\alpha},
\end{eqnarray}
Such interaction possesses the same point group symmetry as the crystal, yet it can yield to symmetry-breaking of the type obtained from quadrupolar interactions \cite{2004-Khavkine-Chung-Oganesyan-Kee}. A mean-field theory entirely analogous to the one mentioned above yields the following instability equation for Fermi surface deformations with $d_{x^2-y^2}$ symmetry:
\begin{equation}
  1=\frac{V}{\pi^2t^*}\int_{-1}^{1-|\mu^*|/2t^*} d\xi \frac{2\xi^2+\xi|\mu^*|/2t^*}{\sqrt{1-\xi^2}
  \sqrt{1-\left(\xi+|\mu^*|/2t^*\right)^2}},
\end{equation}
where $t^*$, $\mu^*$ include a non symmetry-breaking renormalisation of the band structure. This admits a closed-form solution that does not fit in this short note but which is represented in Fig.~\ref{fig.1}~(b). Note (i) that the system is very unstable near half-filling, just as for an explicitly symmetry-breaking interaction \cite{2004-Khavkine-Chung-Oganesyan-Kee}; (ii) that the off-site interaction is necessary, at the mean-field level, for the symmetry-breaking instability ---similar to what happens in the continuum models \cite{2006-Quintanilla-Schofield}.

On a crystal lattice, Friedel oscillations around an impurity can also be used as a probe of symmetry breaking \cite{2007-Doh-Kee}. Coupling of the bare hopping integral $t$ to lattice distortions \cite{2000-Yamase-Kohno} provides another probe. After a $d_{x^2-y^2}$ PI, the ground state energy of the electrons is
\begin{equation}
  \left\langle\hat{H}\right\rangle
  =
  \sum_{{\bf k},\sigma}
  \left[
    -2\tilde{t}_x \cos(k_xa)
    -2\tilde{t}_y \cos(k_ya)
    -\tilde{\mu}
  \right]
\left\langle
    \hat{c}_{{\bf k},\sigma}^{\dagger}
    \hat{c}_{{\bf k},\sigma}
  \right\rangle
\end{equation}
where $\tilde{\mu}\equiv\mu+(\mu^*-\mu)/2$, $\tilde{t}_{x,y}\equiv t+(t_{x,y}^*-t)/2$, and  \(\left\langle
    \hat{c}_{{\bf k},\sigma}^{\dagger}
    \hat{c}_{{\bf k},\sigma}
  \right\rangle 
  =
  \Theta\left[
    -2t_x^* \cos(k_xa)
    -2t_y^* \cos(k_ya)
    -\mu^*
  \right] .\) 
Assuming a linear coupling of $t$ to small variations, $\delta_x a, \delta_y a$,
of the lattice constants, $t \to t - \eta \delta_{x,y} a,$ and minimising the
energy with respect to them (after adding an elastic contribution $\kappa^{-1} (\delta_xa^2+\delta_ya^2)$) we find:
\begin{equation}
  \delta_xa-\delta_ya 
  = 
  \frac{\eta}{\kappa^{-1}}\sum_k 
  \left[
    \cos(k_x a)-\cos(k_ya)
  \right]
  \left\langle
    \hat{c}_{{\bf k},\sigma}^{\dagger}
    \hat{c}_{{\bf k},\sigma}
  \right\rangle.
\end{equation}
Evidently the Pomeranchuk distortion induces some lattice anisotropy. \cite{Powell-Quintanilla}

\section*{Acknowledgements}
JQ acknowledges an Atlas fellowship awarded by CCLRC (now STFC) in association with St. Catherine's College, University of Oxford. CH acknowledges financial support from the Scottish Universities Physics Alliance (SUPA). BJP was supported by the ARC. We thank S. Ramos, D. Cabra and N. I. Gidopoulos for useful discussions.


\begin{thebibliography}{99}

\bibitem{1958-Pomeranchuk}I. Ia. Pomeranchuk, Sov. Phys. JETP {\bf 35}, 524 (1958).
\bibitem{2007-Fradkin-Kivelson-Oganesyan}E. Fradkin, S. A. Kivelson and V. Oganesyan, {\it Science} {\bf 315}, 196 (2007).
\bibitem{2001-Oganesyan-Kivelson-Fradkin}V. Oganesyan, S. A. Kivelson and E. Fradkin, {\it Phys. Rev. B} {\bf 64}, 195109 (2001).
\bibitem{2004-Khavkine-Chung-Oganesyan-Kee}I. Khavkine, Chung-Ho Chung, V. Oganesyan and Hae-Young Kee, {\it Phys. Rev. B} {\bf 70}, 155110 (2004).
\bibitem{2006-Quintanilla-Schofield} J. Quintanilla and A. J. Schofield, {\it Phys. Rev. B} {\bf
74}, 115126 (2006).
\bibitem{2007-Doan-Manousakis}Q. M. Doan and E. Manousakis, Phys. Rev. B {\bf 75}, 195433 (2007).
\bibitem{2006-Varma-Zhu}C. M. Varma and Lijun Zhu, {\it Phys. Rev. Lett.} {\bf 96}, 036405 (2006).
\bibitem{Sr327}S. A. Grigera {\it et al.}, {\it Science} {\bf 306}, 1154 (2004); R. A. Borzi {\it et al.}, {\it Science} {\bf 315}, 214 (2007).
\bibitem{2003-Fiete-Heller}G. A. Fiete and E. J. Heller, {\it Rev. Mod. Phys.} {\bf 75}, 933 (2003).
\bibitem{Schofield-Hooley-Quintanilla}A. J. Schofield, C. Hooley and J. Quintanilla, unpublished.
\bibitem{2001-Valenzuela-Vozmediano}B. Valenzuela and M.A.H. Vozmediano, Phys. Rev. B {\bf 63}, 153103 (2001).
\bibitem{2007-Doh-Kee}Hyeonkin Doh and Hae-Young Kee, Phys. Rev. B {\bf 75}, 233102 (2007).
\bibitem{2000-Yamase-Kohno}H. Yamase and H. Kohno, J. Phys. Soc. Japan {\bf 69}, 2151 (2000).
\bibitem{Powell-Quintanilla}B. J. Powell and J. Quintanilla, unpublished.
\end{thebibliography}
\end{document}